\documentstyle[preprint,eqsecnum,aps,epsf]{revtex}

\begin{document}
\draft
\preprint{SAGA-HE-130-98}
\title{Modification of parton distributions in nuclei}
\author{S. Kumano and K. Umekawa \cite{byline}}
\address{Department of Physics, Saga University \\
         Saga 840-8502, Japan}
%\date{\today}
\date{March 16, 1998}
\maketitle
\begin{abstract}
Nuclear parton distributions are studied in a parton model
with rescaling and recombination mechanisms. Parton $x$
distributions are first calculated at $Q^2$=1 GeV$^2$ by the model,
and they are evolved to larger $Q^2$ in order to be compared with
various $F_2$ data. The experimental shadowing is explained by the
parton-recombination effects. Then, the modification of sea-quark
and gluon distributions is investigated. We show $x$ and $Q^2$
dependent results of the ratios $S_A/S_D$ and $G_A/G_D$. 
They indicate significant shadowing at small $x$ and
large ratios in the medium-$x$ region.
Our theoretical results should be tested by future experiments,
for example, at RHIC (Relativistic Heavy Ion Collider).
Furthermore, our studies are important for finding a signature
of the quark-gluon plasma in heavy-ion reactions.
We also indicate that nuclear effects in the deuteron should
be taken into account for finding the accurate $F_2$ structure
function of the neutron.
\end{abstract}

\narrowtext

%%%%%%%%%%%%%%%%%%%%%%%%%%%%%%%%%%%%%%%%%%%%%%%%%%%%%%%%%%%%%%%%%%%%%%%%%%%%%%%%
\section{Introduction}

Although unpolarized parton distributions in the nucleon are now known
in detail from very small $x$ to relatively large $x$, the distributions
in nuclei are not investigated to such a great extent \cite{summary}.
It was usually assumed in the 1970's that the nuclear structure functions
$F_2$ were simply given by the one for the nucleon multiplied by
the mass number for isoscalar nuclei: $F_2^A=A F_2^N$. 
Nuclear binding energies are very small compared with energy
scales in deep inelastic scattering, so that they were expected to be
insignificant in discussing the structure functions. It is, therefore,
rather startling to find significant modification of the nuclear $F_2$ at
medium $x$ in the European Muon Collaboration (EMC) data \cite{emc83}
in 1983. The nuclear modification is now named the (``old") EMC effect
after the experimental collaboration name.

In the middle of 1980's, theoretical models were proposed for explaining
the nuclear modification in the medium-$x$ region. The first idea after
the EMC finding is a $Q^2$ rescaling model \cite{rescale} in 1983. 
The original motivation for the rescaling model was to consider
that the nucleon size is modified in the nuclear medium.
Because the structure function $F_2^A$ reflects lightcone momentum
distributions of nuclear partons, the size change results
in the modification of $F_2$. Although such modification may exist,
the rescaling model is no more interpreted in the original form.
It is better to be considered as an effective model \cite{rescale87},
which could be consistent with the following binding model.
A conservative idea in terms of nuclear binding was, on the other hand,
proposed in 1985 \cite{binding}. The nuclear $F_2$ is given by
the nucleon's $F_2$ convoluted with a spectral function, which
indicates momentum distributions of the nucleons in the nucleus.
Due to the binding energy $\varepsilon$, the $x$ distribution of $F_2$
is shifted a few percent ($\sim \varepsilon/ m_{_N}$).
Although the magnitude itself is not large, the shift becomes 10$-$20
\% effects in the $F_2$ at medium $x$.
Because the model is successful in explaining a significant portion
of the experimental data, the binding model is now accepted
as the most probable interpretation of the EMC effect at medium $x$.
The model can also explain the large-$x$ region, where the ratio
$F_2^A/F_2^D$ increases as $x\rightarrow 1$, in terms of the
nucleon Fermi motion. We discuss the compatibility of the rescaling
model with the binding model in section \ref{model}.

The small-$x$ region was studied particularly from the late 1980's.
Accurate ratios $F_2^A/F_2^D$ were obtained
experimentally at small $x$ for several nuclei by the New Muon
Collaboration (NMC) \cite{nmcf2} and the Fermilab-E665 collaboration
\cite{e665f2}. It became possible to test theoretical calculations
in comparison with the data. The nuclear modification at small $x$
is called shadowing, which means that internal constituents
are shadowed due to the existence of nuclear surface
constituents. A laboratory frame picture for describing
the shadowing is a vector-meson-dominance (VMD) type model \cite{vmd}.
A virtual photon transforms into vector meson states, 
which then interact with a target nucleus.
The propagation length of the hadronic fluctuation
exceeds the average nucleon separation (2 fm) in nuclei at $x<0.1$,
so that the shadowing occurs because of multiple scattering
of the hadron inside the nucleus.
On the other hand, the shadowing phenomena could be explained
by a parton-recombination model \cite{recomb,cqr,sk,yl}, which
is a description in an infinite momentum frame.
The recombination occurs because the localization size of
a parton with momentum fraction $x$ exceeds 2 fm at $x<0.1$.

We showed in the parton model with the rescaling
and recombination mechanisms that the nuclear $F_2$
could be explained from very small $x$ to large $x$ \cite{sk,nuclq2}.
Although the nuclear $F_2$ structure functions are established,
the sea-quark and gluon distributions are not well tested \cite{ekr}.
Of course, because the $F_2$ at small $x$ 
is essentially given by the sea-quark distribution, 
the sea modification should be equivalent to the $F_2$ modification
in the leading order (LO) of $\alpha_s$.
However, we would rather have independent confirmation, for example,
by the Drell-Yan experiment. Although there exist the Fermilab-E772
Drell-Yan data \cite{e772}, the small-$x$ shadowing part and 
the medium-$x$ region were not probed.
Furthermore, there is little experimental information on the nuclear
gluon distributions \cite{skglue,glue}.
An item of good news is that the Relativistic Heavy Ion Collider (RHIC)
will be completed in the near future, so that the shadowing in the
sea-quark and gluon distributions could be observed experimentally
by the Drell-Yan and direct-photon processes.
Considering these experimental possibilities, we should try to
predict the shadowing behavior theoretically in order to test our
theoretical understanding on high-energy nuclear structure.

The major purpose of this paper is extend the parton model in Ref. \cite{sk}
to the studies of the sea-quark and gluon distributions in nuclei.
In addition, the parton distributions in the nucleon are
updated to the new ones. The small-$x$ distributions were particularly
modified due to the HERA $F_2$ data after the publication of Ref. \cite{sk}.
Another purpose is to study the parton distributions in the deuteron. 
They were assumed to be the same as those of
the nucleon in the previous publications \cite{sk,skglue}.
In section \ref{model}, our parton model is introduced with explanation
for the $Q^2$ rescaling and parton-recombination mechanisms.
In section \ref{f2}, the $F_2$ structure functions are calculated
for various nuclei in order to be compared with measured experimental data.
In section \ref{sea-glue}, the model is extended to the sea-quark
and gluon distributions. Their $x$ and $Q^2$ dependent results are shown.
As a byproduct of our investigation, we can discuss the relation between
the deuteron and neutron structure functions.
In section \ref{deuteron}, our results are compared
with the experimental ratios $F_2^n/F_2^p$.
The summary is given in section \ref{sum}.

%%%%%%%%%%%%%%%%%%%%%%%%%%%%%%%%%%%%%%%%%%%%%%%%%%%%%%%%%%%%%%%%%%%%%%%%%%%%%%%%
%%%%%%%%%%%%%%%%%%%%%%%%%%%%%%%%%%%%%%%%%%%%%%%%%%%%%%%%%%%%%%%%%%%%%%%%%%%%%%%%
\section{Nuclear model}\label{model}

People studied possible physics mechanisms for interpreting
the medium-$x$ modification of the $F_2$ structure functions
just after the EMC finding. A few years later, the small-$x$
mechanisms were investigated independently from the medium-$x$
physics. It is rather unfortunate that enough efforts are not
made to unify these pictures to produce $F_2$ from small $x$
to large $x$. For example, the parton distributions in the wide $x$
range are necessary for studying $Q^2$ dependence of the structure
functions because they should be supplied as input distributions.
For this reason, we set up a parton model which could describe the
parton distributions in the whole $x$ region in a dynamically consistent
way \cite{sk}. We calculated the modification of the nuclear parton
distributions at certain $Q^2$ ($\equiv Q_0^{\, 2}$) by the rescaling
and recombination mechanisms. Then, they are evolved to larger $Q^2$
where experimental data are taken. 
Because the same model is used in our studies, we explain only the
essential part in the following. The interested reader may look
at the original papers \cite{rescale,rescale87,cqr,sk}
for finding the detailed formalism.

%%%%%%%%%%%%%%%%%%%%%%%%%%%%%%%%%%%%%%%%%%%%%%%%%%%%%%%%%%%%%%%%%%%%%%%%%%%%%%%%
\subsection{$Q^2$ rescaling}\label{rescaling}

As explained in the introduction, the $Q^2$ rescaling model was originally
proposed as a first explicit quark signature in nuclear physics in terms
of nucleon-size modification in the nuclear medium \cite{rescale}.
As a result, the nuclear $F_2$ structure function is related to
the nucleon's $F_2$ by a simple $Q^2$ rescaling,
\begin{equation}
F_2^A(x,Q^2) = F_2^N (x, \xi_{_A} Q^2)
\ \ \ {\rm with} \ \
\xi_{_A} = ( \lambda_A^{\, 2} / \lambda_N^{\, 2} )
          ^{\alpha_s(\mu_A^{\, 2} )/\alpha_s(Q^2)}
\ ,
\label{eqn:rescale}
\end{equation}
where $\xi_{_A}$ is called rescaling parameter. The factors $\lambda_A$
and $\lambda_N$ are confinement radii for the nucleus A and the nucleon,
respectively. 

The proponents of this model modified the interpretation \cite{rescale87}.
They rather consider the rescaling model as an effective model,
which could include the binding and/or nucleon modification effects.
In order to justify this idea, they used factorization-scale independence.
It means that physics observables should not depend on the scale which is
introduced artificially by hand. This is generally used in structure functions,
which are separated into nonperturbative quantities (parton distributions)
and perturbative ones (coefficient functions) by the separation scale $\mu$.
The scale $\mu$ is introduced artificially, so that the structure functions
should be independent of the scale. It leads to so called renormalization group
equation. In our situation, the factorization means that 
a nuclear structure function is separated into a nucleon momentum distribution
and the corresponding structure function in the nucleon. The factorization
scale independence is not, to be exact, proved in this case. However, 
we consider it rather as a working hypothesis: the observables, nuclear
structure functions, are independent of the artificial separation into
nuclear physics and subnucleon physics. Then, the nuclear binding model
could be also expressed in term of the $Q^2$ rescaling terminology.
For example, the binding energy and Fermi momentum in the Fermi gas model
are related to the rescaling parameter by \cite{rescale87}
$\overline{\varepsilon} / m_N = - \gamma_2^{NS} \kappa_A / (2 \beta_0)$
and
$k_F^2 / m_N^2 = 5 (2 \gamma_2^{NS} - \gamma_3^{NS}) \kappa_A /(2 \beta_0) 
  +O({\overline\varepsilon}^2 / m_N^2)$,
where
$ 1 - \kappa_A = \alpha_s (\xi_A \, Q^2) / \alpha_s (Q^2) $.
Therefore, the rescaling model could be considered as an effective model
which could include the binding-type nuclear effects and possibly also
subnucleon-type contributions. In this sense, the rescaling parameter
$\xi_A$ could be determined by fitting some experimental data, rather
than calculating it in a particular nuclear model such as the one in
the rescaling paper of 1985 \cite{rescale}. In any case, the original
fits are no longer valid if Fermi-motion effects are included \cite{bm}.
As explained in section \ref{f2}, the parameter for the calcium nucleus
$\xi_{Ca}$ is determined by the $F_2^{Ca}/F_2^D$ data at medium $x$. 
The other nuclear values are calculated by using
the A dependence relation of the rescaling model.

%%%%%%%%%%%%%%%%%%%%%%%%%%%%%%%%%%%%%%%%%%%%%%%%%%%%%%%%%%%%%%%%%%%%%%%%%%%%%%%%
\subsection{Parton recombination}\label{recomb}

We use an infinite momentum picture, the parton-recombination model
\cite{recomb,cqr,sk,yl}, for describing the small-$x$ shadowing physics,
although there are other descriptions, for example, VMD-type models
\cite{vmd} and Pomeron models \cite{pomeron}. These different models
explain the nuclear $F_2$ structure functions in some ways
or other, so the appropriate description may be determined by studying
other observables such as nuclear gluon distributions or possibly the
valence-quark distributions at small $x$ \cite{val}.

The parton recombination, which means the interaction of partons in different
nucleons, is used as a model for interpreting the nuclear shadowing
\cite{recomb,cqr,sk,yl}. This mechanism could occur by the following reason.
In an infinite momentum frame, the average longitudinal nucleon 
separation in a Lorentz contracted nucleus is
$L \approx (2.2~ fm) M_A /P_A
     = (2.2~ fm) m_{_N}/p_{_N}$,
and the longitudinal localization size of a parton with momentum $xp_{_N}$
is $\Delta L \approx 1/(xp_{_N})$.
At small $x$ ($x<0.1$), the localization size becomes smaller than
the average nucleon separation. It means that partons from different
nucleons could interact significantly. The interaction is called
parton recombination or parton fusion.
According to this mechanism, the modification of a parton distribution
$p_3(x_3)$ is given by
\begin{equation}
\Delta p_3 (x_3) = K \int dx_1 dx_2 \, p_1 (x_1) \, p_2 (x_2) \, 
             \Gamma_{p_1 p_2 \rightarrow p_3} (x_1,x_2,x_3) \, 
             \delta (x_3 - x_1 - x_2)
\ ,
\label{eqn:recomb}
\end{equation}
for the recombination process $p_1 p_2 \rightarrow p_3$.
Here, the factor $K$ is defined by
$K=9A^{1/3}\alpha_s/(2R_0^2 Q^2)$ with the expression for
the nuclear radius $R=R_0 A^{1/3}$. 
The function $\Gamma_{p_1 p_2 \rightarrow p_3} (x_1,x_2,x_3)$
is called parton fusion function. Because the recombination
is opposite to the splitting process, the $\Gamma_{p_1 p_2 \rightarrow p_3}$
is related to the corresponding splitting function $P_{p_1 \leftarrow p_3}$
in the Dokshitzer-Gribov-Lipatov-Altarelli-Parisi (DGLAP) $Q^2$ evolution
equations:
\begin{equation}
\Gamma_{p_1 p_2 \rightarrow p_3} (x_1,x_2,x_3) =
     {\frac{x_1 x_2}{x_3^2}} \, 
     P_{p_1 \leftarrow p_3} (x_1 / x_3) \,
     C_{p_1 p_2 \rightarrow p_3}
\ ,
\end{equation}
where $C_{p_1 p_2 \rightarrow p_3}$ is the color-factor ratio, for example
$\displaystyle{
C_{qG \rightarrow q}
=\sum_{(l,a),k}^{\textstyle{-}} (t_{kl}^a)^* t_{kl}^a
/\sum_{(k),l,a}^{\textstyle{-}} (t_{lk}^a)^* t_{lk}^a
}$. The averages are taken over the initial color indices,
for example, $(k)$ indicates the average over the index $k$.
There is also a momentum-cutoff factor $w(x)=exp(-m_{_N}^2 z_0^2 x^2/2)$
for the leak-out partons. It indicates that large-momentum partons
are confined in the small region so that they do not leak out from
the nucleon. It is known that $z_0$=2 fm is an appropriate value
\cite{lle,sk}, so that it is used throughout our analysis.
Because the recombination formalism itself is discussed in detail in
Refs. \cite{cqr,sk}, we do not repeat it here.
The explicit equations are given in the Appendix part of Ref. \cite{sk}.

%%%%%%%%%%%%%%%%%%%%%%%%%%%%%%%%%%%%%%%%%%%%%%%%%%%%%%%%%%%%%%%%%%%%%%%%%%%%%%%%
%%%%%%%%%%%%%%%%%%%%%%%%%%%%%%%%%%%%%%%%%%%%%%%%%%%%%%%%%%%%%%%%%%%%%%%%%%%%%%%%
\section{Structure functions F$_2^A$}\label{f2}

In order to calculate nuclear parton distributions, it is necessary to
employ a set of parton distributions in the nucleon.
The optimum parton distributions are obtained by fitting many experimental
data from various processes. There are three major groups in the studies of
the unpolarized parametrization. They are
CTEQ (Coordinated Theoretical/Experimental Project on QCD 
      Phenomenology and Tests of the Standard Model),
GRV (Gl\"uck, Reya, and Vogt), and MRS (Martin, Roberts, and Stirling).
Among these parametrizations, we decided to take the MRS-R2 distributions
\cite{mrs-r} without any serious reason. Of course, the details of our
numerical results depend on the input distributions; however, the essential
results should not be altered. For simplicity, the MRS-R2 is modified
to have the flavor symmetric sea, $\bar u=\bar d=\bar s=Sea/6$,
in the following analyses.

We have to determine the point $Q_0^{\, 2}$, at which the initial
nuclear distributions are calculated. It is considered as a parameter
in our model, and it is determined in the following way.
In order to explain the large shadowing, for example in
$F_2^{Ca}/F_2^D$, it is necessary to take rather small $Q_0^{\, 2}$
because the recombination effect has a higher-twist nature
($\sim 1/Q_0^{\, 2}$). On the other hand, perturbative Quantum
Chromodynamics (QCD) is assumed in evolving the calculated distributions
at $Q_0^2$ to those at experimental $Q^2$ points. Therefore, $Q_0^{\, 2}$
should be in the perturbative region. With these two reasons, we end up
taking $Q_0^{\, 2}\sim$1 GeV$^2$.
Because the MRS-R2 distributions are set up at $Q^2$=1 GeV$^2$,
it is simply taken as the MRS-R2 point, namely $Q_0^{\, 2}$=1 GeV$^2$,
in the following analyses.

Using the MRS-R2 distributions, we calculate first the rescaling effects
for the valence quark distributions at $Q_0^{\, 2}=$1 GeV$^2$. 
Because this procedure violates the momentum conservation, the gluon
distribution is multiplied by a constant amount so as to satisfy
the conservation. Then, the obtained distributions are used as
input distributions for calculating the recombination effects.
The experimental $Q^2$ values are different depending on the $x$ region:
they are typically a few GeV$^2$ at small $x$ and a few dozen GeV$^2$
at large $x$. In order to compare with the experimental data, we evolve
the obtained nuclear distributions at $Q^2$=1 GeV$^2$ to those
at $Q^2$=5 GeV$^2$ by the parton-recombination (PR) $Q^2$ evolution
equations \cite{glr,mq}. In particular, we use the equations proposed
by Mueller and Qiu \cite{mq} and their numerical solution in Ref. \cite{bf1}.
The QCD scale parameter of the MRS-R2 is $\Lambda$=0.344 GeV.

In calculating the rescaling, it is necessary to find the rescaling
parameter $\xi_A$. As we mentioned in the previous section,
$\xi_{Ca}$ is determined for explaining the experimental medium-$x$ ratios,
$F_2^{Ca}/F_2^D$, reasonably well.
It is found that $\xi_{Ca}$=1.74 is the appropriate
value at $Q^2$=1 GeV$^2$. The rescaling parameters of other nuclei
are determined by the relationship of the original rescaling model.
Namely, the scale parameter is determined by
$\mu_A =  (\lambda_{N}/\lambda_A) \mu_{N}$ with
$\mu_{N}^2$=0.66 GeV$^2$. The effective confinement size is
given by $\lambda_{Ca}/\lambda_N$=1.194 according to
Eq. (\ref{eqn:rescale}) and $\xi_{Ca}$=1.74.
The confinement size is 5.0 percent larger than the one
($\lambda_{Ca}/\lambda_N$=1.137) in Ref. \cite{rescale}.
Therefore, other nuclear confinement radii in Ref. \cite{rescale}
are also multiplied by the factor 1.050. 
To be explicit, they are given for the deuteron, helium,
carbon, and tin as
$\lambda_D/\lambda_N   $=1.066, $\xi_D   $=1.186,
$\lambda_{He}/\lambda_N$=1.133, $\xi_{He}$=1.437,
$\lambda_C/\lambda_N   $=1.159, $\xi_C   $=1.557,
$\lambda_{Sn}/\lambda_N$=1.235, and $\xi_{Sn}$=2.001,
where we assumed that the nuclei are
$^4$He, $^{12}$C, $^{40}$Ca, and $^{118}$Sn in the theoretical
calculations. All the rescaling parameters are obtained
at $Q^2$=1 GeV$^2$. Using these parameters,
we calculate the rescaling modification. 

In the previous publication \cite{sk}, the deuteron structure
function $F_2^D$ is assumed to be equal to the one for the nucleon
in calculating the ratios $F_2^A/F_2^D$. A special attention should be paid
for the deuteron if we wish to calculate it in the recombination model.
There is an explicit nuclear radius factor $R_0$ in Eq. (\ref{eqn:recomb})
and also in the PR evolution equations.
However, it is well known that the deuteron radius
does not follow the rule, $R=R_0 A^{1/3}$ with $R_0$=1.1 fm. 
We estimate it by the relation $R=\sqrt{5<r^2>/12}$, where
$r$ is the distance between the proton and the neutron.
Then, the constant $R_0$ becomes $R_0$=1.56 fm for the deuteron.
It is fairly large in comparison with the usual nuclear value $R_0$=1.1 fm
because the deuteron is a loosely bound system.
It means that the recombination effects in the deuteron
are much smaller than those in other nuclei. Throughout this paper,
we use $R_0$=1.56 fm for the deuteron.

\begin{figure}
\vspace{0.5cm}
\hspace{0.0cm}
\epsfxsize=12.0cm
\centering{\epsfbox{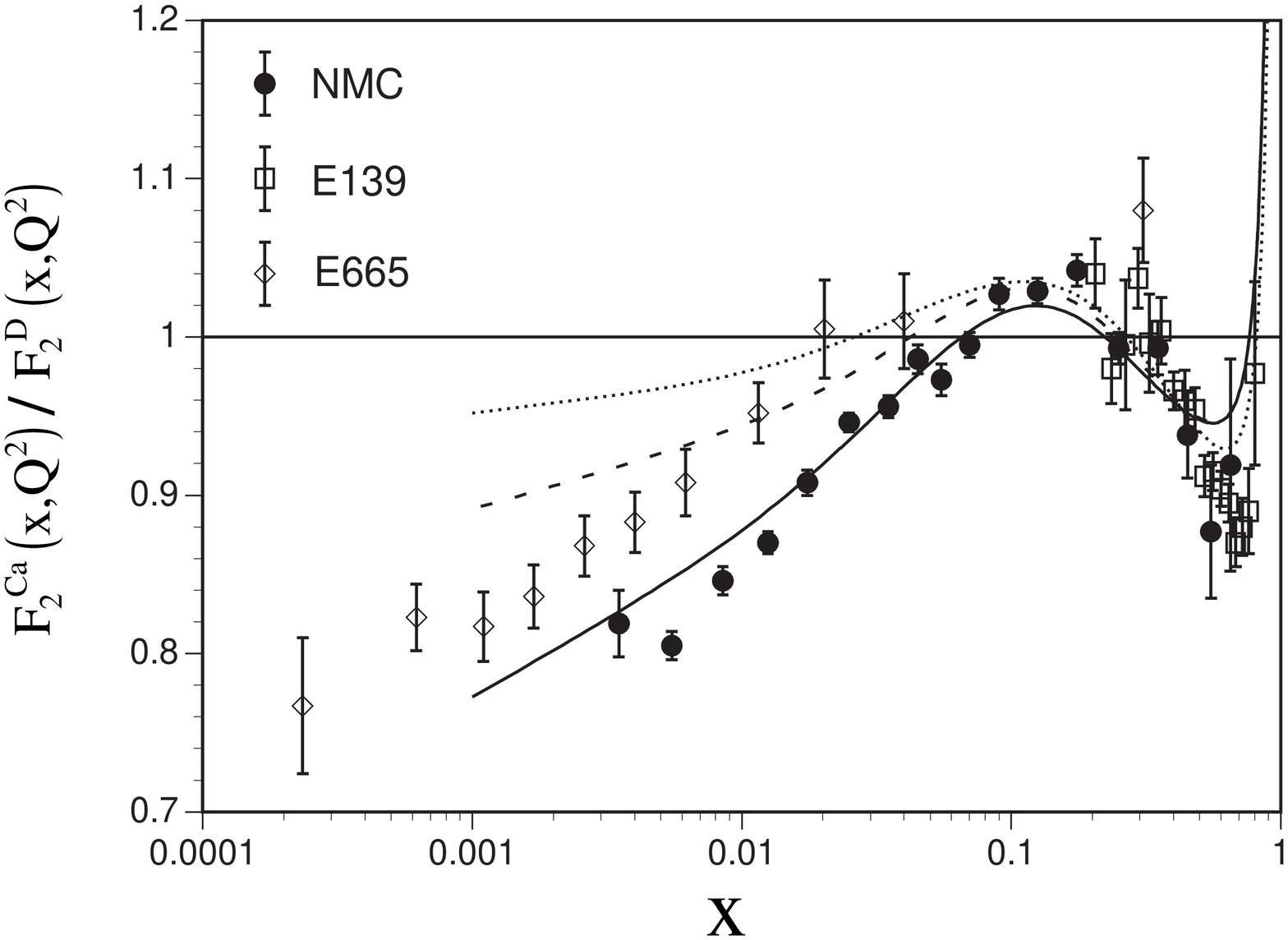}}
\vspace{0.0cm}
\caption{Theoretical results for the ratio $F_2^{Ca}/F_2^D$ are
         compared with the NMC, Fermilab-E665, and SLAC-E139 data. 
         The dotted curve is the initial ratio at $Q^2$=1 GeV$^2$.
         The distributions are evolved to $Q^2$=5 GeV$^2$, and
         the dashed and solid curves are obtained by the DGLAP and
         PR evolution equations, respectively.}
\label{fig:f2ca}
\end{figure}

\begin{figure}
\vspace{0.0cm}
\hspace{0.0cm}
\epsfxsize=12.0cm
\centering{\epsfbox{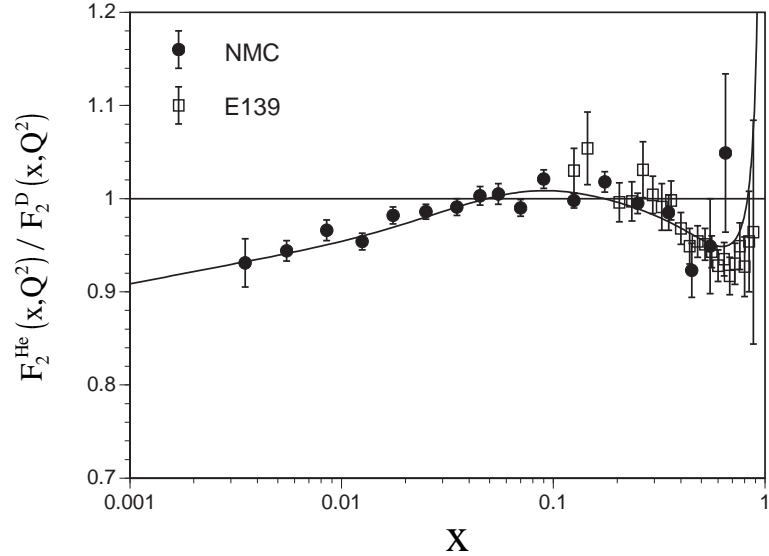}}
\vspace{0.0cm}
\caption{Nuclear modification of $F_2$ is shown for the helium nucleus.
         The theoretical result is shown by the solid curve
         at $Q^2$=5 GeV$^2$.}
\label{fig:f2he}
\end{figure}

\begin{figure}
\vspace{0.0cm}
\hspace{0.0cm}
\epsfxsize=12.0cm
\centering{\epsfbox{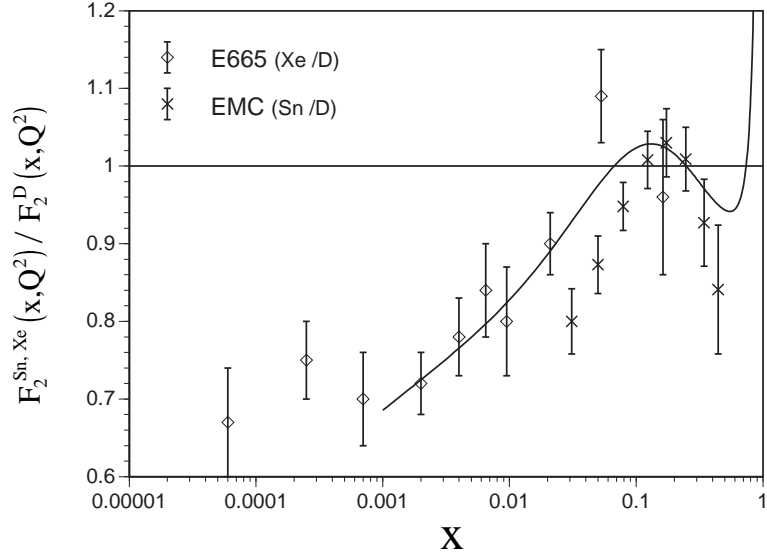}}
\vspace{0.0cm}
\caption{The modification is calculated for the tin nucleus.
         The theoretical result is shown by the solid curve
         at $Q^2$=5 GeV$^2$.
         The E665 data are also shown even though they are taken
         for a different nucleus, xenon.}
\label{fig:f2sn}
\end{figure}

With these preparations, we are ready for calculating the $F_2$
ratios $F_2^A/F_2^D$. First, we show the results for the calcium
nucleus in Fig. \ref{fig:f2ca}. The dotted curve indicates the distribution
at $Q^2$=1 GeV$^2$. It should be noted that the convolution integrals
of the next-to-leading-order (NLO) coefficient functions with the parton
distributions are calculated for obtaining the $F_2$.
The ratio is evolved to the ones at $Q^2$=5 GeV$^2$.
The dashed curve is obtained by using the NLO DGLAP evolution equations,
and the solid one is by the PR evolution equations.
The $Q^2$ dependence from $Q^2$=5 GeV$^2$
to larger values is rather small as we show it later.
The experimental data are those of the NMC \cite{nmcf2},
the Fermilab-E665 \cite{e665f2}, and the SLAC-E139 \cite{slacf2}.
The theoretical curves are not shown at $x<0.001$ because the $Q^2$
values of the E665 data are very small and our perturbative calculations
cannot be compared with them.
We note that there are significant differences between the NMC and
E665 data. From Fig. \ref{fig:f2ca}, we find that our theoretical shadowing
is rather small at $Q^2$=1 GeV$^2$ in comparison with the experimental one.
The evolved distribution by the DGLAP shows also smaller shadowing.
The NMC data can be explained only if the PR evolution equations are used. 
It is also interesting to find that the ratio $F_2^{Ca}/F_2^D >>1$
at large $x$ is explained as a recombination
effect without the explicit Fermi-motion corrections.
We apply the model for other nuclei. 
Our results are compared with experimental data for a small
and large nuclei, helium and tin, in  Figs. \ref{fig:f2he} and
\ref{fig:f2sn} where the EMC data \cite{emcf2} are also shown. 
The solid curves indicate our theoretical results at $Q^2$=5 GeV$^2$
with the PR evolution. 
Although the modification tends to be small at $x\approx 0.7$,
we find in these figures that our model can explain the
$x$ dependence reasonably well in the wide $x$ region ($0.001<x<0.8$)
for the small, medium, and large size nuclei.

\begin{figure}
\vspace{0.0cm}
\hspace{0.0cm}
\epsfxsize=12.0cm
\centering{\epsfbox{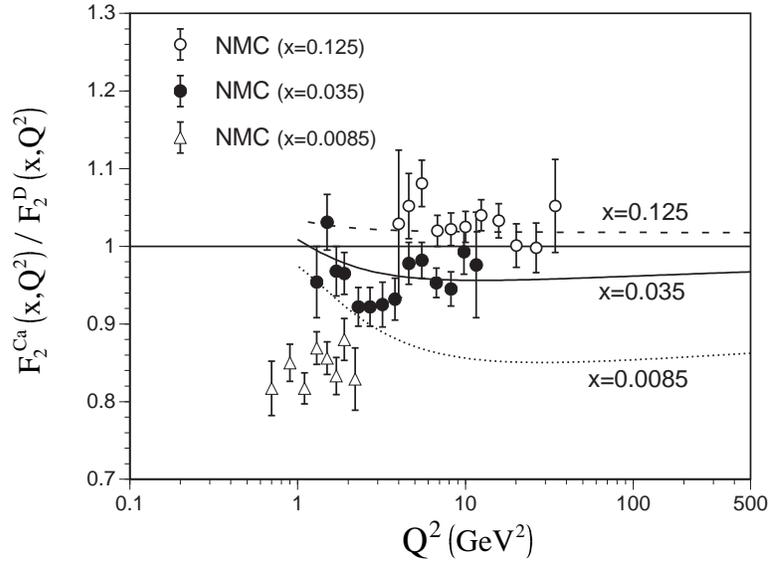}}
\vspace{0.0cm}
\caption{Calculated $Q^2$ dependence of the ratio $F_2^{Ca}/F_2^D$ is
         compared with the NMC data. The dotted, solid, and dashed
         curves are obtained at $x$=0.0085, 0.035, and 0.125,
         respectively.}
\label{fig:f2q2}
\end{figure}

Next, $Q^2$ dependence of $F_2^{Ca}/F_2^D$ is calculated at the fixed
$x$ points, $x$=0.0085, 0.035, and 0.125. Then, the theoretical ratios
are compared with the NMC data in Fig. \ref{fig:f2q2}.
The calculated results are shown by the dotted,
solid, and dashed curves for $x$=0.0085, 0.035, and 0.125, respectively.
The figure indicates that our model is successful in explaining
the $Q^2$ dependence at $x$=0.035 and 0.12; however, it is not
very successful in the small $Q^2$ region at $x$=0.0085.
Namely, the calculated shadowing is too small to explain the
experimental one in the region $Q^2\sim$1 GeV$^2$.
Therefore, the model needs to be studied further whether
it is in fact impossible to explain the shadowing behavior
at small $Q^2$. We leave this problem as our future research topic.

We found that our model is successful in explaining the gross properties
of the nuclear structure functions $F_2^A$. It is interesting to extend
the model to the studies of sea-quark and gluon distributions in nuclei.
This topic is discussed in the next section.

\vfill\eject
%%%%%%%%%%%%%%%%%%%%%%%%%%%%%%%%%%%%%%%%%%%%%%%%%%%%%%%%%%%%%%%%%%%%%%%%%%%%%%%%
%%%%%%%%%%%%%%%%%%%%%%%%%%%%%%%%%%%%%%%%%%%%%%%%%%%%%%%%%%%%%%%%%%%%%%%%%%%%%%%%
\section{Sea-quark and gluon distributions}\label{sea-glue}

\begin{figure}
\vspace{0.0cm}
\hspace{0.0cm}
\epsfxsize=12.0cm
\centering{\epsfbox{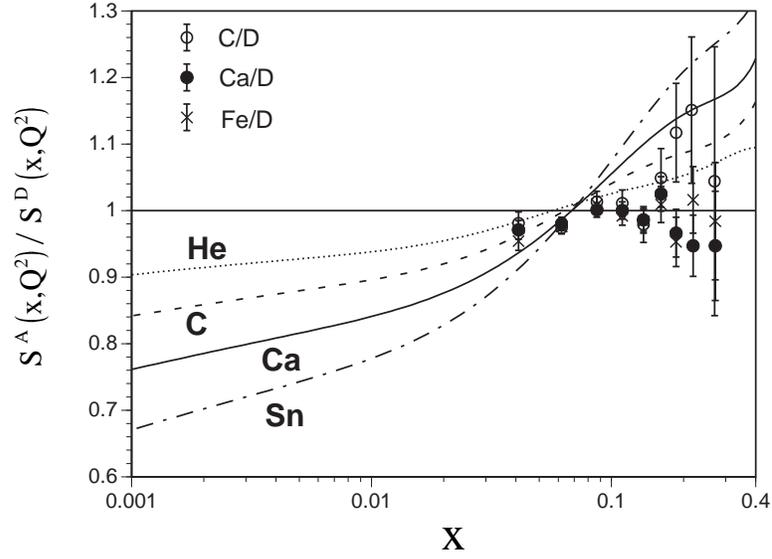}}
\vspace{0.0cm}
\caption{Calculated $x$ dependence of the sea-quark modification is
         shown by the dotted, dashed, solid, and dot-dashed curves
         for the ratios $S^{He}/S^D$, $S^{C}/S^D$, $S^{Ca}/S^D$, and
         $S^{Sn}/S^D$ at $Q^2$=5 GeV$^2$. They are compared with
         the Fermilab-E772 data for the carbon, calcium, and iron.}
\label{fig:sx}
\end{figure}

The nuclear modification of sea-quark and gluon distributions is not
well investigated. Because the $F_2$ at small $x$ is determined by
the sea-quark distributions, the sea-quark shadowing is almost
equal to the $F_2$ shadowing. 
However, there is no independent experimental confirmation
of the sea shadowing. There are Fermilab-E772 Drell-Yan data \cite{e772},
which measured the ratios, $S^{C}/S^D$, $S^{Ca}/S^D$, and $S^{Fe}/S^D$
in the $x$ range $0.04<x<0.27$. The data are shown in Fig. \ref{fig:sx}.
Although the iron data are often quoted in suggesting that there is no
nuclear modification in the measured $x$ range, it is not very obvious
by noting the $A$ dependence and the experimental accuracy. The nuclear
sea-quark distributions have been studied extensively in the context
of the pion-excess model \cite{piexcess}. However, we do not step into
the pion-excess problem in this paper.

\begin{figure}
\vspace{0.0cm}
\hspace{0.0cm}
\epsfxsize=12.0cm
\centering{\epsfbox{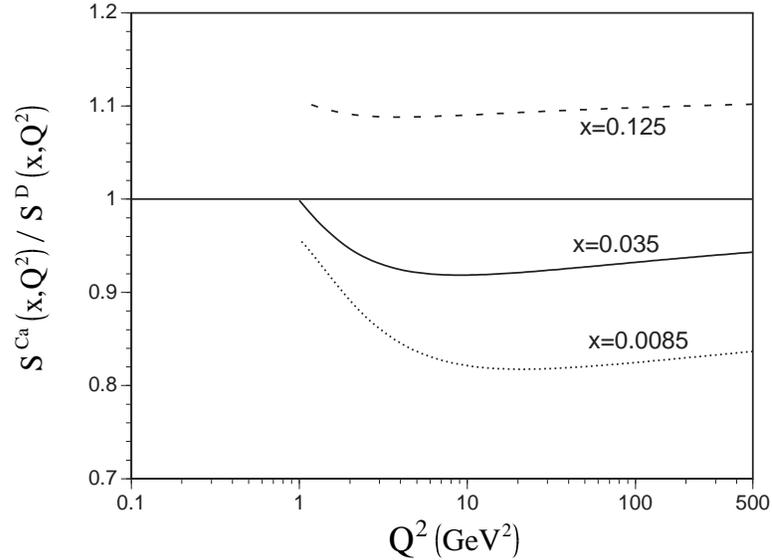}}
\vspace{0.0cm}
\caption{Calculated $Q^2$ dependence of the sea-quark modification is
         shown for the calcium-deuteron ratio $S^{Ca}/S^D$.
         The dotted, solid, and dashed curves are obtained
         at $x$=0.0085, 0.035, and 0.125, respectively.}
\label{fig:sq2}
\end{figure}

We show our theoretical results for the helium, carbon, calcium, and tin nuclei
in Fig. \ref{fig:sx}. The calculated sea-quark shadowing is not exactly
equal to the one for $F_2$ because the gluon distribution also contributes
to $F_2$ through the coefficient function. However, the gluon corrections
are small so that both shadowing results are almost the same.
The ratio becomes large at medium $x$ due to the recombination effects.
The data for $F_2^A/F_2^D$ in Figs. \ref{fig:f2ca} and \ref{fig:f2he}
indicate antishadowing features at $x \approx 0.1$.
Our model explains this region by the large sea-quark ratios $S^A/S^D$
in Fig. \ref{fig:sx}. However, we are annoyed that our ratios are
too large to explain the experimental sea-quark distributions
in the region $0.1<x<0.2$.
If there were no nuclear modification as suggested by the E772 data,
the $F_2$ antishadowing should be interpreted as the valence-quark
antishadowing. On the other hand, there seems no physics mechanism
to produce the valence antishadowing in such a $x$ region \cite{val}.
Of course, it is possible to produce the antishadowing artificially
if the baryon-number conservation is imposed with the valence-quark
shadowing and the medium-$x$ EMC effect \cite{fsl}. However, it does
not solve the physics problem. We hope that experimentalists measure
the sea-quark ratios from very small $x$ to medium $x$ accurately
for several different nuclei. The accurate data should be able to
clarify the theoretical problems. At this stage, there is no
experimental information on the $Q^2$ dependence of the sea-quark
ratios, of course, except for the one implied by the $F_2$ data at small $x$.  
We show the ratios for the calcium nucleus at
$x$=0.0085, 0.035, and 0.125 in Fig. \ref{fig:sq2}.
The curves in the figure are almost the same as those in
Fig. \ref{fig:f2q2} except at the $x$=0.125 results where
the valence-quark distributions also contribute.

The detailed sea-quark distributions will be investigated experimentally
at RHIC (Relativistic Heavy Ion Collider).
We may wait a few years for the accurate data.
In addition, flavor distributions of the sea could be an interesting
topic. For example, the $\bar u/\bar d$ asymmetry in nuclei is important
for testing a dynamical aspect of the nuclear parton model \cite{sksu2}.
There is also a possibility to measure the nuclear $\bar u/\bar d$ asymmetry
at Fermilab \cite{nuclub-db}.

Next, the nuclear gluon distributions are calculated.
We have already shown the first version in Ref. \cite{skglue};
however, the results were not obtained in a consistent manner with
the $F_2$ structure functions. 
Here, the model was tested first by comparing various $F_2^A/F_2^D$ data.
We found that it is a fairly successful model in section \ref{f2},
so that it is reasonable to extend it to the studies of
the sea-quark and gluon distributions. 
To find the accurate nuclear gluon distributions is important 
not only for establishing the nuclear model at high energies but also
for other applications to heavy-ion physics.
For example, the J/$\psi$ suppression may be interpreted as
a quark-gluon plasma signature; however, the initial gluon
distributions in heavy nuclei are not known. It would be,
at least partially, a local gluon shadowing effect \cite{local}. 
In order to find the correct interpretation, the information on the precise
gluon distributions is inevitable. At this stage, most people just assume
that nuclear gluon distributions are equal to the nucleon's distribution.
However, it is known that the recombination mechanism produces
strong gluon shadowing \cite{skglue,glue}.
Despite their importance, it is unfortunate that the gluon distributions
are not actually measured. There are implicit data
on the nuclear modification and explicit data taken by the NMC
\cite{nmcglue} for the ratio $G^{Sn}/G^C$. However, the data
are not accurate enough to enlighten us.

\begin{figure}
\vspace{0.0cm}
\hspace{0.0cm}
\epsfxsize=12.0cm
\centering{\epsfbox{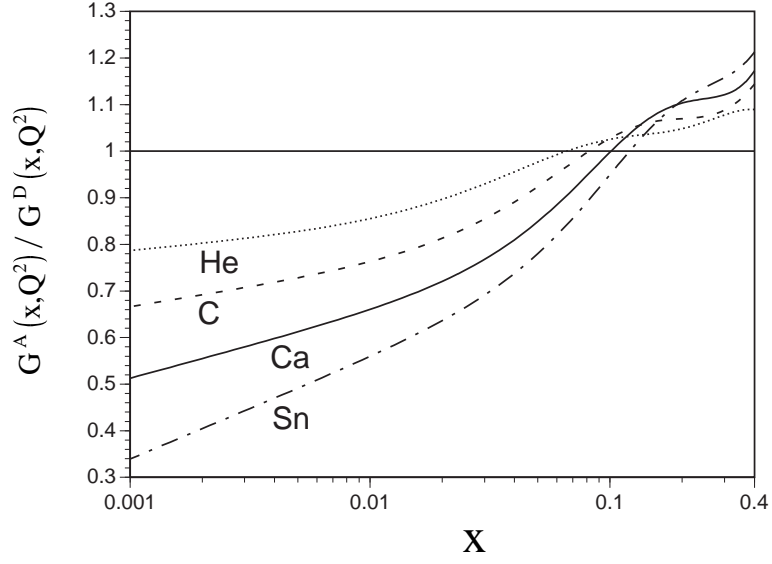}}
\vspace{0.0cm}
\caption{Calculated $x$ dependence of the gluon modification is shown
         by the dotted, dashed, solid, and dot-dashed curves for
         the ratios $G^{He}/G^D$, $G^{C}/G^D$, $G^{Ca}/G^D$,
         and $G^{Sn}/G^D$ at $Q^2$=5 GeV$^2$.}
\label{fig:gx}
\end{figure}

\begin{figure}
\vspace{0.0cm}
\hspace{0.0cm}
\epsfxsize=12.0cm
\centering{\epsfbox{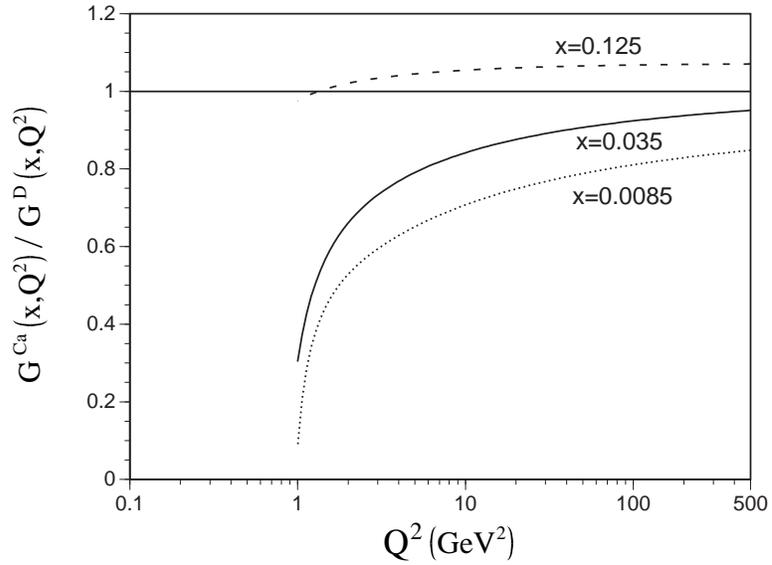}}
\vspace{0.0cm}
\caption{Calculated $Q^2$ dependence of gluon modification is shown
         for the calcium-deuteron ratio $G^{Ca}/G^D$.
         The dotted, solid, and dashed
         curves are obtained at $x$=0.0085, 0.035, and 0.125.}
\label{fig:gq2}
\end{figure}

We show the calculated nuclear modification in Figs. \ref{fig:gx}
and \ref{fig:gq2}. Figure \ref{fig:gx} shows the $x$ dependence
for the helium, carbon, calcium, and tin nuclei.
The figure indicates that the gluon shadowing is significantly
larger than that of quarks, and the ratio becomes large at medium $x$
due to the recombination effects.
The $Q^2$ dependence in Fig. \ref{fig:gq2} shows that the gluon
shadowing is very large at small $Q^2$ and it has large $Q^2$
dependence at small $Q^2$.

It is interesting to test these predictions by future experiments. 
The RHIC facility is appropriate for finding the gluon shadowing
\cite{a-glue}, for example, through the direct-photon production process.
In order to find accurate sea-quark and gluon distributions,
experimental information is crucial.

%%%%%%%%%%%%%%%%%%%%%%%%%%%%%%%%%%%%%%%%%%%%%%%%%%%%%%%%%%%%%%%%%%%%%%%%%%%%%%%%
%%%%%%%%%%%%%%%%%%%%%%%%%%%%%%%%%%%%%%%%%%%%%%%%%%%%%%%%%%%%%%%%%%%%%%%%%%%%%%%%
\section{Deuteron}\label{deuteron}

The nuclear corrections have been calculated in the deuteron
for obtaining the structure-function ratios $F_2^A/F_2^D$
in section \ref{f2}. Here, we discuss the correction effects on a sum rule. 
Because there is no fixed target for the neutron, the deuteron
or $^3 He$ is usually used for measuring neutron structure functions.
The nuclear corrections are important in extracting
the ``neutron" structure functions from the deuteron
and $^3 He$ data.
For example, the Gottfried sum rule is given by the difference
between $F_2^p$ and $F_2^n$ \cite{skpr}:
\begin{equation}
I_G \, = \, \int_0^1 \frac{dx}{x} \, 
                    [ F_2^p(x,Q^2) - F_2^n(x,Q^2) ] 
\, = \, \frac{1}{3} \, + \, \frac{2}{3} \int_0^1 dx \, 
[\bar u(x,Q^2) -\bar d(x,Q^2)]
\ .
\label{eqn:gott}
\end{equation}
The NMC used the measured proton and deuteron structure functions,
$F_2^p$ and $F_2^D$, for evaluating the difference \cite{nmcgott},
\begin{equation}
F_2^p -F_2^n= 2 \, F_2^D \, \frac{1-F_2^n/F_2^p}{1+F_2^n/F_2^p}
\ , \ \ \ F_2^n/F_2^p=2F_2^D/F_2^p-1
\ .
\end{equation}
These equations, of course, assume no nuclear correction.
As it was discussed in section \ref{f2}, the nuclear structure
functions are modified in the whole $x$ region.
In particular, it is known that the small-$x$ part, namely
the deuteron shadowing, is important
for obtaining the ``correct" sum in Eq. (\ref{eqn:gott}).

\begin{figure}
\vspace{0.0cm}
\hspace{0.0cm}
\epsfxsize=12.0cm
\centering{\epsfbox{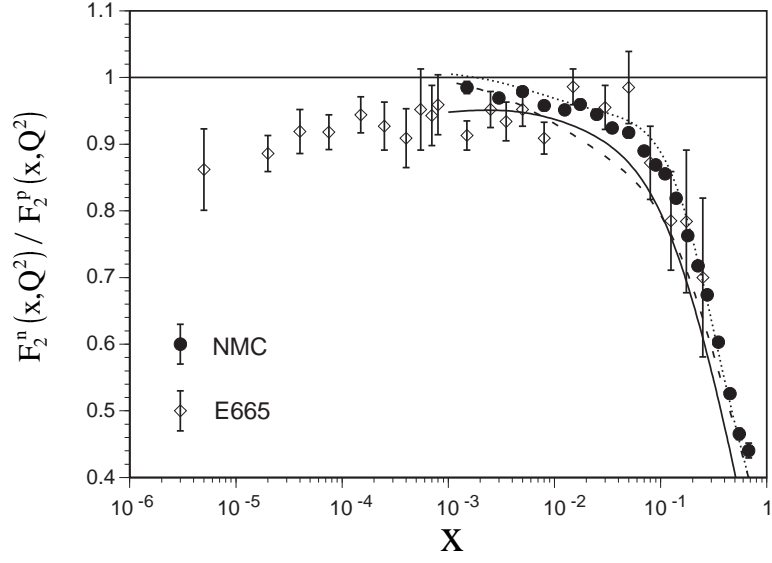}}
\vspace{0.0cm}
\caption{The ratio of the neutron and proton structure functions
         is shown. The dotted curve is the MRS-R2
         at $Q^2$=1 GeV$^2$. The flavor symmetric one is shown
         by the dashed curve at $Q^2$=1 GeV$^2$. The solid curve
         is the evolution result at $Q^2$=5 GeV$^2$,
         and it is calculated by $2F_2^D/F_2^p-1$
         with the nuclear corrections in the deuteron.}
\label{fig:f2n-p}
\end{figure}

The $F_2^n/F_2^p$ ratios are measured by various groups.
We compare our theoretical results with the data by the NMC \cite{nmcn-p}
and the Fermilab-E665 \cite{e665f2n-p} in Fig. \ref{fig:f2n-p}.
The dotted curve is the MRS-R2 distribution ($\bar u \ne \bar d$)
for the $F_2^n/F_2^p$ ratio at $Q^2$=1 GeV$^2$,
and the dashed curve is the flavor symmetric one which is given by
the modified MRS-R2 so as to have $\bar u = \bar d =\bar s = Sea/6$
at $Q^2$=1 GeV$^2$. 
The calculated distribution $2F_2^D/F_2^p-1$ at $Q^2$=5 GeV$^2$
is shown by the solid curve, which includes the nuclear corrections.
Because the E665 small-$x$ data are taken at very small $Q^2$
where the perturbative QCD would not work, the theoretical
curves are not plotted at small $x$ ($<$0.001).

Because the MRS-R2 was created so as to explain the NMC data
as well as many other data, the dotted curve is in agreement with
the NMC ratios. The flavor symmetric distribution, which is shown
by the dashed curve, is well below the NMC data at small $x$ where
sea-quark contributions are significant. The solid curve includes
the nuclear modification due to the rescaling and recombinations,
and it also contains $Q^2$ evolution effects from $Q^2$=1 GeV$^2$ to
5 GeV$^2$. As a result, the solid curve is below the NMC data in
the $x$ region ($0.0015<x<0.675$).
It is mainly because our model used the flavor symmetric input
in order to see the nuclear modification effects.
However, the difference is partly attributed to the incomplete
analyses by the NMC and MRS-R2 in the sense that the nuclear corrections
are completely neglected. In order to illustrate the significance of
the corrections, the nuclear contribution to the Gottfried sum is
estimated by 
\begin{equation}
I_G=I_G^{\, 0} + \int_{x_{min}=0}^1 \frac{dx}{x} \, 
                       2 \, [ \, F_2^D (x) - F_2^N (x) \, ]
\ .
\end{equation}
The $I_G^{\, 0}$ is the sum without the nuclear correction,
and the second term ($\equiv \delta I_G$) is the correction.
The integral is evaluated for the structure functions at $Q^2$=5 GeV$^2$,
and we obtain $\delta I_G=-0.013$ for the integral
from $x_{min}=0.004$. It is the same order of magnitude as the other recent
shadowing estimates \cite{d-gott}. This kind of correction should be
taking into account in discussing the Gottfried sum rule and also
other sum rules with neutron structure functions.

\vfill\eject
%%%%%%%%%%%%%%%%%%%%%%%%%%%%%%%%%%%%%%%%%%%%%%%%%%%%%%%%%%%%%%%%%%%%%%%%%%%%%%%%
%%%%%%%%%%%%%%%%%%%%%%%%%%%%%%%%%%%%%%%%%%%%%%%%%%%%%%%%%%%%%%%%%%%%%%%%%%%%%%%%
\section{Summary}\label{sum}

We have investigated the parton model with the $Q^2$ rescaling
and parton-recombination mechanisms for describing the parton
distributions in nuclei. The nuclear modification of the $x$
distributions is calculated at $Q^2$=1 GeV$^2$. 
If they are evolved to larger $Q^2$, for example
$Q^2$=5 GeV$^2$, by the parton-recombination evolution equations,
they become comparable with the experimental $F_2$ shadowing.

The model was then extended to the studies of sea-quark and gluon
distributions in nuclei. We showed the $x$ and $Q^2$ dependent
results of the ratios $S_A/S_D$ and $G_A/G_D$.
They indicate large nuclear shadowing at small $x$ and large ratios
in the medium-$x$ region. These predictions could be tested by
the experiments at RHIC. The understanding of nuclear parton
distributions is very important for heavy-ion physics studies,
for example, in finding a signature of the quark-gluon plasma.

We also indicated in our model that the nuclear effects should be taking
into account properly in order to discuss the sum rules with neutron
structure functions, which are extracted from the deuteron data.

\vspace{0.7cm}
%%%%%%%%%%%%%%%%%%%%%%%%%%%%%%%%%%%%%%%%%%%%%%%%%%%%%%%%%%%%%%%%%%%%%%%%%%%%%%
%%%%%%%%%%%%%%%%%%%%%%%%%%%%%%%%%%%%%%%%%%%%%%%%%%%%%%%%%%%%%%%%%%%%%%%%%%%%%%
\section*{{\bf Acknowledgment}}
\addcontentsline{toc}{section}{\protect\numberline{\S}{Acknowledgments}}

SK and KU thank M. Miyama for discussions on nuclear structure functions
and for his help in their numerical analyses.

%%%%%%%%%%%%%%%%%%%%%%%%%%%%%%%%%%%%%%%%%%%%%%%%%%%%%%%%%%%%%%%%%%%%%%%%%%%%%%%%
%%%%%%%%%%%%%%%%%%%%%%%%%%%%%%%%%%%%%%%%%%%%%%%%%%%%%%%%%%%%%%%%%%%%%%%%%%%%%%%%

%%%%%%%%%%%%%%%%%%%%%%%%%%%%%%%%%%%%%%%%%%%%%%%%%%%%%%%%%%%%%%%%%%%%%%%%%%%%%%%%
%%%%%%%%%%%%%%%%%%%%%%%%%%%%%%%%%%%%%%%%%%%%%%%%%%%%%%%%%%%%%%%%%%%%%%%%%%%%%%%%

%%%%%%%%%%%%%%%%%%%%%%%%%%%  figure 1 %%%%%%%%%%%%%%%%%%%%%%%%%%%%%%%%%%
%%%%%%%%%%%%%%%%%%%%%%%%%%%  figure 1 %%%%%%%%%%%%%%%%%%%%%%%%%%%%%%%%%%

\vspace{0.0cm}
%%%%%%%%%%%%%%%%%%%%%%%%%%%  figure 2 %%%%%%%%%%%%%%%%%%%%%%%%%%%%%%%%%%
%%%%%%%%%%%%%%%%%%%%%%%%%%%  figure 2 %%%%%%%%%%%%%%%%%%%%%%%%%%%%%%%%%%

%%%%%%%%%%%%%%%%%%%%%%%%%%%%%%%%%%%%%%%%%%%%%%%%%%%%%%%%%%%%%%%%%%%%%%%%%%%%%%%%
\vfill\eject

$\ \ \ $

\vspace{0.0cm}
%%%%%%%%%%%%%%%%%%%%%%%%%%%  figure 3 %%%%%%%%%%%%%%%%%%%%%%%%%%%%%%%%%%
%%%%%%%%%%%%%%%%%%%%%%%%%%%  figure 3 %%%%%%%%%%%%%%%%%%%%%%%%%%%%%%%%%%

\vspace{1.0cm}
%%%%%%%%%%%%%%%%%%%%%%%%%%%  figure 4 %%%%%%%%%%%%%%%%%%%%%%%%%%%%%%%%%%
%%%%%%%%%%%%%%%%%%%%%%%%%%%  figure 4 %%%%%%%%%%%%%%%%%%%%%%%%%%%%%%%%%%

%%%%%%%%%%%%%%%%%%%%%%%%%%%%%%%%%%%%%%%%%%%%%%%%%%%%%%%%%%%%%%%%%%%%%%%%%%%%%%%%
\vfill\eject

$\ \ \ $

\vspace{-0.5cm}
%%%%%%%%%%%%%%%%%%%%%%%%%%%  figure 5 %%%%%%%%%%%%%%%%%%%%%%%%%%%%%%%%%%
%%%%%%%%%%%%%%%%%%%%%%%%%%%  figure 5 %%%%%%%%%%%%%%%%%%%%%%%%%%%%%%%%%%

\vspace{1.0cm}
%%%%%%%%%%%%%%%%%%%%%%%%%%%  figure 6 %%%%%%%%%%%%%%%%%%%%%%%%%%%%%%%%%%
%%%%%%%%%%%%%%%%%%%%%%%%%%%  figure 6 %%%%%%%%%%%%%%%%%%%%%%%%%%%%%%%%%%

%%%%%%%%%%%%%%%%%%%%%%%%%%%%%%%%%%%%%%%%%%%%%%%%%%%%%%%%%%%%%%%%%%%%%%%%%%%%%%%%
\vfill\eject

$\ \ \ $

\vspace{0.0cm}
%%%%%%%%%%%%%%%%%%%%%%%%%%%  figure 7 %%%%%%%%%%%%%%%%%%%%%%%%%%%%%%%%%%
%%%%%%%%%%%%%%%%%%%%%%%%%%%  figure 7 %%%%%%%%%%%%%%%%%%%%%%%%%%%%%%%%%%

\vspace{1.0cm}
%%%%%%%%%%%%%%%%%%%%%%%%%%%  figure 8 %%%%%%%%%%%%%%%%%%%%%%%%%%%%%%%%%%
%%%%%%%%%%%%%%%%%%%%%%%%%%%  figure 8 %%%%%%%%%%%%%%%%%%%%%%%%%%%%%%%%%%

%%%%%%%%%%%%%%%%%%%%%%%%%%%%%%%%%%%%%%%%%%%%%%%%%%%%%%%%%%%%%%%%%%%%%%%%%%%%%%%%
\vfill\eject

$\ \ \ $

\vspace{0.0cm}
%%%%%%%%%%%%%%%%%%%%%%%%%%%  figure 9 %%%%%%%%%%%%%%%%%%%%%%%%%%%%%%%%%%
%%%%%%%%%%%%%%%%%%%%%%%%%%%  figure 9 %%%%%%%%%%%%%%%%%%%%%%%%%%%%%%%%%%

\end{document}